\documentclass{ws-procs9x6}
\usepackage{amscd,graphics,axodraw}

\newcommand{\cc}{\mathrm{cc}}
\newcommand{\comp}{\mathrm{comp}}
\newcommand{\es}{\raisebox{-0.3cm}{$\emptyset$}}
\newcommand{\gggg}{\mathfrak{g}}
\newcommand{\la}{\lambda}
\newcommand{\lb}{\overline{\ell}}
\newcommand{\nt}{\nu^\bullet}
\newcommand{\ntt}{\tilde{\nu}^\bullet}

\newcommand{\bin}[2]{\left( \begin{array}{c} #1\\ #2 \end{array} \right)}
\newcommand{\qbin}[2]{\left[ \begin{array}{c} #1\\ #2 \end{array} \right]}
\newcommand{\rk}{\mathrm{rk}}
\newcommand{\sbf}{\vec{s}}
\newcommand{\sigbf}{\vec{\sigma}}

\newcommand{\rig}[1]{\makebox[0.3cm]{#1}}
\newcommand{\su}{\mathfrak{su}}
\newcommand{\tr}{\mathrm{tr}}
\newcommand{\veps}{\varepsilon}
\newcommand{\vphi}{\varphi}
\newcommand{\wt}{\mathrm{wt}}
\newcommand{\Bt}{\tilde{B}}
\newcommand{\Complex}{\mathbb{C}}
\newcommand{\Conf}{\mathrm{C}}

\newcommand{\Hil}{\mathcal{H}}
\newcommand{\Jt}{J^\bullet}
\newcommand{\Jtt}{\tilde{J}^\bullet}
\newcommand{\La}{\Lambda}
\newcommand{\Phit}{\tilde{\Phi}}
\newcommand{\R}{\mathbb{R}}
\newcommand{\RC}{\mathrm{RC}}
\newcommand{\Path}{\mathcal{P}}
\newcommand{\Z}{\mathbb{Z}}
\newcommand{\Zn}{\Z_{\ge 0}}

\begin{document}

\pagestyle{plain}

\title{Rigged configurations and the Bethe Ansatz}

\author{Anne Schilling}
\address{Department of Mathematics,\\
 University of California,\\
 One Shields Avenue,\\
 Davis, CA 95616-8633, U.S.A.\\
 E-mail: anne@math.ucdavis.edu}

\maketitle

\abstracts{
This note is a review of rigged configurations and the Bethe Ansatz.
In the first part, we focus on the algebraic Bethe Ansatz for the 
spin 1/2 XXX model and explain how rigged configurations label the 
solutions of the Bethe equations. This yields the bijection between rigged
configurations and crystal paths/Young tableaux of Kerov, Kirillov and
Reshetikhin. In the second part, we discuss a generalization of this 
bijection for the symmetry algebra $D_n^{(1)}$, based on work in 
collaboration with Okado and Shimozono.
}

\section{Introduction}

These notes arose from three lectures presented
at the Summer School on Theoretical Physics "Symmetry and Structural 
Properties of Condensed Matter" held in Myczkowce, Poland, 
on September 11-18, 2002.
We review the algebraic Bethe Ansatz in the simple
setting of the spin 1/2 XXX model, explain the physical meaning
of rigged configurations and give a bijection between rigged 
configurations and crystal bases for type $D_n^{(1)}$ \cite{OSS:2002}
generalizing the bijection of Kerov, Kirillov and 
Reshetikhin \cite{KKR:1986,KR:1988} for type $A_n^{(1)}$.

The Bethe Ansatz originated in a paper by Bethe \cite{Bethe:1931} in
1931 in which he studied the eigenvectors and eigenfunctions of
the Hamiltonian of the Heisenberg antiferromagnet. The method he used 
is today often called the coordinate Bethe Ansatz, distinguishing it
from the algebraic Bethe Ansatz that will be presented here.
The algebraic Bethe Ansatz is a generalization of the coordinate
Bethe Ansatz and is one of the most important outcomes of the
quantum inverse scattering method introduced 
in~\cite{Sklyanin:1982,SF:1978,STF:1980}. 
The quantum inverse scattering method has unified
the treatment of quantum integrable systems considering each model
as a representation of the quantum monodromy matrix which satisfies
certain commutation relations. The algebraic Bethe Ansatz is based
on the idea of constructing eigenvectors of the Hamiltonian
(resp. trace of the monodromy matrix) by creation and annihilation
operators on a vacuum; the elements of the monodromy matrix
play the role of these operators. The eigenvectors are parametrized
by solutions of a system of algebraic equations, called the Bethe
equations. The solutions in turn are labeled by combinatorial
objects called rigged configurations.

We consider $\gggg$-invariant models where $\gggg$ is the symmetry
algebra. The Hilbert
space is the tensor product of irreducible representations of $\gggg$
denoted $\Hil=h_1\otimes \cdots \otimes h_N$. The Bethe vectors are the highest
weight vectors in the decomposition into irreducible components of
$\Hil$. It is known from representation theory that the highest weight
vectors are also labeled by Young tableaux (see for example \cite{Fulton:1997})
or certain paths in crystal theory 
(see for example~\cite{HKOTT:2001,HKOTY:1999,Nakashima:1993}).
Assuming the completeness of the Bethe vectors, this suggests a bijection
between rigged configurations and Young tableaux/crystal paths.
For $\gggg=\mathfrak{gl}_n$ such a bijection was given by Kirillov and
Reshetikhin \cite{KR:1988} and generalized in \cite{KSS:2002}.

Analogous bijections for all $\gggg$ of nonexceptional affine type
were recently proven in \cite{OSS:2002} for tensor products of the
fundamental representation. An important property of all these
bijections is that they preserve statistics that can be defined
on the set of rigged configurations and paths, respectively.
As a corollary it follows that one-dimensional configuration sums
defined in terms of crystal paths have fermionic formulas.
Fermionic formulas reflect the quasiparticle structure of the
underlying model and also reveal the statistics of the quasiparticles.
For general affine Kac-Moody algebras fermionic formulas were conjectured
by Hatayama, Kuniba, Okado, Takagi, Tsuboi and 
Yamada~\cite{HKOTT:2001,HKOTY:1999}. For type $A_n^{(1)}$ they were
proven in \cite{KSS:2002} and for nonexceptional types in special
cases in \cite{OSS:2002}.

The paper is organized as follows. In section \ref{sec:model}
we review the algebraic Bethe Ansatz for the spin 1/2 XXX model
and derive the Bethe equations. In section \ref{sec:sol} we present
the solutions of the Bethe equations parametrized by rigged configurations
and discuss the bijection between rigged configurations and paths
in section \ref{sec:rig}. Sections \ref{sec:model} and \ref{sec:sol}
follow the presentation of Faddeev \cite{F:1998}.
In sections \ref{sec:general}-\ref{sec:bij} the bijection between paths
and rigged configurations is generalized to types $A_n^{(1)}$ and
$D_n^{(1)}$ based on work in collaboration with Okado and Shimozono
\cite{OSS:2002}. Crystal bases are introduced in section \ref{sec:crystals}
and section \ref{sec:rc} states the fermionic formula and rigged configurations
in the generalized set-up. The bijection is given explicitly in 
section \ref{sec:bij}.

\section{Bethe Ansatz for the XXX model}
\label{sec:model}

In this section we discuss the algebraic Bethe Ansatz
for the example of the spin $1/2$ XXX Heisenberg chain.
This is a one-dimensional quantum spin chain on $N$ sites
with periodic boundary conditions. It is defined on the Hilbert space
$\Hil_N = \bigotimes_{n=1}^N h_n$ where in this case $h_n=\Complex^2$
for all $n$. Associated to each site is a local spin variable
$\sbf = \frac{1}{2} \sigbf$ where
\begin{equation*}
\sigbf=(\sigma^1,\sigma^2,\sigma^3)=
\left( \left(\begin{array}{cc} 0&1\\ 1&0\end{array}\right),
\left( \begin{array}{cc} 0&-i\\ i&0 \end{array} \right),
\left( \begin{array}{cc} 1&0\\ 0&-1 \end{array} \right) \right)
\end{equation*}
are the Pauli matrices.
The spin variable acting on the $n$-th site is given by
\begin{equation*}
\sbf_n = I \otimes \cdots \otimes I \otimes \sbf \otimes I
 \otimes \cdots \otimes I
\end{equation*}
where $I$ is the identity operator and $\sbf$ is in the $n$-th
tensor factor. We impose periodic boundary conditions
$\sbf_n = \sbf_{n+N}$.

The Hamiltonian of the spin $1/2$ XXX model is
\begin{equation*}
H_N = J \sum_{n=1}^N \left(\sbf_n \cdot \sbf_{n+1} - \frac{1}{4}\right).
\end{equation*}
Our goal is to determine the eigenvectors and eigenvalues of 
$H_N$ in the antiferromagnetic regime $J>0$ in the limit when
$N\to \infty$.

The main tool will be the Lax operator $L_{n,a}(\la)$, also called the local
transition matrix. It acts on $h_n\otimes \Complex^2$ where $\Complex^2$ is an
auxiliary space and is defined as
\begin{equation*}
L_{n,a}(\la) = \la I_n \otimes I_a + i \sbf_n \otimes \sigbf_a.
\end{equation*}
Here $I_n$ and $I_a$ are unit operators acting on $h_n$ and the
auxiliary space $\Complex^2$, respectively; $\la$ is a complex parameter,
called the spectral parameter. Writing the action on the auxiliary space as
a $2\times 2$ matrix, we have
\begin{equation}\label{eq:L aux}
L_{n}(\la) = \left( \begin{array}{cc} \la + i s_n^3 & i s_n^-\\
i s_n^+ & \la - i s_n^3 \end{array} \right)
\end{equation}
where $s_n^\pm = s_n^1 \pm i s_n^2$.

The crucial fact is that the Lax operator satisfies commutation relations
in the auxiliary space $V=\Complex^2$. Altogether there are 16 relations
which can be written compactly in tensor notation.
Given two Lax operators $L_{n,a_1}(\la)$ and $L_{n,a_2}(\mu)$ defined
in the same quantum space $h_n$, but different auxiliary spaces
$V_1$ and $V_2$, the products $L_{n,a_1}(\la)L_{n,a_2}(\mu)$ and
$L_{n,a_2}(\mu)L_{n,a_1}(\la)$ are defined on the triple tensor
product $h_n\otimes V_1\otimes V_2$.
There exists an operator $R_{a_1,a_2}(\la-\mu)$ defined on
$V_1\otimes V_2$ such that
\begin{equation}\label{eq:com rel}
R_{a_1,a_2}(\la-\mu) L_{n,a_1}(\la)L_{n,a_2}(\mu)
= L_{n,a_2}(\mu)L_{n,a_1}(\la) R_{a_1,a_2}(\la-\mu).
\end{equation}
Explicitly, the $R$-matrix $R_{a_1,a_2}(\la)$ is given by
\begin{equation*}
R_{a_1,a_2}(\la) = \left( \la +\frac{i}{2}\right) I_{a_1}\otimes I_{a_2}
 + \frac{i}{2} \sigbf_{a_1} \otimes \sigbf_{a_2}.
\end{equation*}
To deduce the 16 relations explicitly, one may write
(\ref{eq:com rel}) as matrices in the auxiliary space 
$V_1\otimes V_2$ using the convention 
$(A\otimes B)^{ij}_{k\ell}=A_{ij}B_{k\ell}$ where
\begin{equation*}
M^{ij}_{k\ell} = \left( \begin{array}{cccc}
 M_{11}^{11} & M_{12}^{11} & M_{11}^{12} & M_{12}^{12}\\
 M_{21}^{11} & M_{22}^{11} & M_{21}^{12} & M_{22}^{12}\\
 M_{11}^{21} & M_{12}^{21} & M_{11}^{22} & M_{12}^{22}\\
 M_{21}^{21} & M_{22}^{21} & M_{21}^{22} & M_{22}^{22}
\end{array} \right).
\end{equation*}
In this notation the $R$-matrix reads
\begin{equation*}
R(\la)= \left( \begin{array}{cccc}
a(\la) &0&0&0\\
0& b(\la)&c(\la)&0\\
0& c(\la)&b(\la)&0\\
0&0&0&a(\la)
\end{array} \right)
\end{equation*}
where $a(\la)=\la+i$, $b(\la)=\la$ and $c(\la)=i$.

Geometrically, the Lax operator $L_{n,a}(\la)$ can be interpreted 
as the transport between sites $n$ and $n+1$ of the quantum
spin chain. Hence
\begin{equation*}
T_{N,a}(\la) = L_{N,a}(\la) \cdots L_{1,a}(\la)
\end{equation*}
is the monodromy around the circle (recall that we assume periodic
boundary conditions). In the auxiliary space write
\begin{equation*}
T_{N}(\la) = \left( \begin{array}{cc}
A(\la) & B(\la)\\
C(\la) & D(\la)
\end{array} \right)
\end{equation*}
with entries in the full Hilbert space $\Hil_N$. From (\ref{eq:com rel})
it is clear that the monodromy matrix satisfies the
following commutation relation
\begin{equation}\label{eq:com rel mon}
R_{a_1,a_2}(\la-\mu)T_{N,a_1}(\la)T_{N,a_2}(\mu)
= T_{N,a_2}(\mu)T_{N,a_1}(\la)R_{a_1,a_2}(\la-\mu).
\end{equation}
Explicitly, some of the relations contained in (\ref{eq:com rel mon})
are
\begin{eqnarray}\label{eq:com}
[B(\la),B(\mu)]&=&0 \nonumber\\
A(\la)B(\mu)&=&f(\la-\mu) B(\mu)A(\la)+g(\la-\mu)B(\la)A(\mu)\\
D(\la)B(\mu)&=&h(\la-\mu)B(\mu)D(\la)+k(\la-\mu)B(\la)D(\mu)\nonumber
\end{eqnarray}
where
\begin{equation*}
\begin{array}{ll}
f(\la)=\frac{\la-i}{\la} & \qquad g(\la)=\frac{i}{\la}\\
h(\la)=\frac{\la+i}{\la} & \qquad k(\la)=-\frac{i}{\la}.
\end{array}
\end{equation*}

It is well-known \cite{F:1998,FT:1981,KBI:1993} that the Hamiltonian
is given in terms of the monodromy matrix as
\begin{equation*}
H_N = \frac{i}{2} \frac{d}{d\la} \ln t_N(\la)|_{\la=i/2}-\frac{N}{2}
\end{equation*}
where $t_N(\la)=\tr T_N(\la) = A(\la)+D(\la)$.

Let $\omega_n=\left( \begin{array}{c} 1\\0\end{array} \right)$. 
In the auxiliary space the Lax operator is triangular on $\omega_n$
\begin{equation*}
L_n(\la)\omega_n = \left( \begin{array}{cc} \la+\frac{i}{2} & *\\
0 & \la-\frac{i}{2} \end{array} \right) \omega_n
\end{equation*}
where $*$ stands for an for us irrelevant quantity. This follows
directly from (\ref{eq:L aux}). On the Hilbert
space $\Hil_N$ we define $\Omega= \bigotimes_n \omega_n$ so that
\begin{equation*}
T_N(\la) \Omega = \left( \begin{array}{cc} \alpha^N(\la) & * \\
0 & \delta^N(\la) \end{array} \right) \Omega
\end{equation*}
where $\alpha(\la)=\la+\frac{i}{2}$ and $\delta(\la)=\la-\frac{i}{2}$.
Equivalently this means that
\begin{eqnarray*}
C(\la) \Omega &=& 0\\
A(\la) \Omega &=& \alpha^N(\la) \Omega\\
D(\la) \Omega &=& \delta^N(\la) \Omega
\end{eqnarray*}
so that $\Omega$ is an eigenstate of $A(\la)$ and $D(\la)$
and hence also of $t_N(\la)=A(\la)+D(\la)$.

The claim is that the other eigenvectors of $t_N(\la)$ are of the form
\begin{equation*}
\Phi(\la,\La) = B(\la_1)\cdots B(\la_n) \Omega
\end{equation*}
where the lambdas $\La=\{\la_1,\ldots,\la_n\}$ 
satisfy a set of algebraic relations, called the
Bethe equations. We will derive these now.

{}From the commutation relations (\ref{eq:com}) we find that
\begin{eqnarray*}
A(\la)B(\la_1)\cdots B(\la_n)\Omega &=& \prod_{k=1}^n f(\la-\la_k)
\alpha^N(\la) B(\la_1)\cdots B(\la_n)\Omega\\
&+&\sum_{k=1}^n M_k(\la,\La) B(\la_1)\cdots \hat{B}(\la_k)\cdots
B(\la_n)B(\la)\Omega.
\end{eqnarray*}
The first term on the right hand side is obtained by using only the
first term on the right hand side of (\ref{eq:com}). The other terms
come from a combination of the application of the first and second term
when moving $A$ past the $B$'s. In general the coefficients $M_k(\la,\La)$
are quite involved using the explicit formulas. However, $M_1(\la,\La)$
is obtained by using the second term in (\ref{eq:com}) moving
$A(\la)$ past $B(\la_1)$ followed by applications of the first
term in (\ref{eq:com}) only. This yields
\begin{equation*}
M_1(\la,\La) = g(\la-\la_1) \prod_{k=2}^n f(\la_1-\la_k) 
 \alpha^N(\la_1).
\end{equation*}
Note that the $B$'s commute with each other by (\ref{eq:com}). Hence
$M_j(\la,\La)$ can be obtained from $M_1(\la,\La)$ by replacing
$\la_1$ by $\la_j$ so that
\begin{equation*}
M_j(\la,\La) = g(\la-\la_j) \prod_{\stackrel{k=1}{k\neq j}}^n 
 f(\la_j-\la_k) \alpha^N(\la_j).
\end{equation*}

Similarly, 
\begin{eqnarray*}
D(\la)B(\la_1)\cdots B(\la_n)\Omega &=& \prod_{k=1}^n h(\la-\la_k)
\delta^N(\la) B(\la_1)\cdots B(\la_n)\Omega\\
&+&\sum_{k=1}^n N_k(\la,\La) B(\la_1)\cdots \hat{B}(\la_k)\cdots
B(\la_n)B(\la)\Omega
\end{eqnarray*}
where
\begin{equation*}
N_j(\la,\La) = k(\la-\la_j) \prod_{\stackrel{k=1}{k\neq j}}^n 
 h(\la_j-\la_k) \delta^N(\la_j).
\end{equation*}

For $\Phi(\la,\La)$ to be an eigenvector of $t_N(\la)=A(\la)+D(\la)$
the terms
\begin{eqnarray*}
&&\sum_{k=1}^n M_k(\la,\La) B(\la_1)\cdots \hat{B}(\la_k)\cdots
B(\la_n)B(\la)\Omega\\
&+&\sum_{k=1}^n N_k(\la,\La) B(\la_1)\cdots \hat{B}(\la_k)\cdots
B(\la_n)B(\la)\Omega
\end{eqnarray*}
need to cancel. Since $g(\la-\la_j)=-k(\la-\la_j)$ this happens if
the set of lambda's $\La$ satisfy the following set of equations
\begin{equation*}
\prod_{\stackrel{k=1}{k\neq j}}^n f(\la_j-\la_k) \alpha^N(\la_j)
=\prod_{\stackrel{k=1}{k\neq j}}^n h(\la_j-\la_k) \delta^N(\la_j)
\end{equation*}
for all $j=1,2,\ldots,n$. Explicitly this reads
\begin{equation}\label{eq:bethe}
\left(\frac{\la_j+\frac{i}{2}}{\la_j-\frac{i}{2}}\right)^N
=\prod_{\stackrel{k=1}{k\neq j}}^n \frac{\la_j-\la_k+i}{\la_j-\la_k-i}
\end{equation}
called the Bethe equations. In this case the eigenvalues of 
$\Phi(\la,\La)$ are
\begin{equation*}
\alpha^N(\la) \prod_{k=1}^n f(\la-\la_k)
+\delta^N(\la) \prod_{k=1}^n h(\la-\la_k).
\end{equation*}
In the next section we will study solutions to (\ref{eq:bethe}) in the 
limit $N\to \infty$.

\section{Solutions to the Bethe equations}
\label{sec:sol}

Let us rewrite (\ref{eq:bethe}) in the following way
\begin{equation}\label{eq:bethe 1}
\left(\frac{\la+\frac{i}{2}}{\la-\frac{i}{2}}\right)^N
=\prod_{\stackrel{\la'\in \La}{\la'\neq \la}} 
 \frac{\la-\la'+i}{\la-\la'-i}
\end{equation}
where $\la\in\La=\{\la_1,\ldots,\la_n\}$.

Suggested by numerical analysis, it is assumed that in the limit
$N\to\infty$ the $\la$'s form strings.
This hypothesis is called the string hypothesis.
A string of length $\ell=2M+1$, where $M$ is an integer or half-integer
depending on the parity of $\ell$, is a set of $\la$'s of the form
\begin{equation*}
\la^M_{jm} = \la^M_j + im
\end{equation*}
where $\la^M_j\in\R$ and $-M\le m\le M$ is integer or half-integer
depending on $M$.
The index $j$ satisfies $1\le j\le m_\ell$ where $m_\ell$
is the number of strings of length $\ell$.
A decomposition of $\{\la_1,\ldots,\la_n\}$ into strings is called
a configuration. Each configuration is parametrized by $\{m_\ell\}$.
It follows that
\begin{equation*}
\sum_\ell \ell m_\ell = n.
\end{equation*}

Now take (\ref{eq:bethe 1}) and multiply over a string
\begin{eqnarray}\label{eq:inter}
\prod_{m=-M}^M &&
 \left( \frac{\la^M_j+i(m+\frac{1}{2})}{\la^M_j+i(m-\frac{1}{2})}
 \right)^N\nonumber\\
&=&\prod_{m=-M}^M \prod_{\stackrel{M',j',m'}{(M',j',m')\neq (M,j,m)}}
\frac{\la^M_j-\la^{M'}_{j'}+i(m-m'+1)}{\la^M_j-\la^{M'}_{j'}+i(m-m'-1)}.
\end{eqnarray}
Many of the terms on the left and right cancel so that this equation
can be rewritten as
\begin{equation}\label{eq:bethe 2}
e^{i N p_M(\la^M_j)} = \prod_{\stackrel{M',j'}{(M',j')\neq (M,j)}} 
 e^{i S_{MM'}(\la_j^M-\la_{j'}^{M'})},
\end{equation}
in terms of the momentum and scattering matrix
\begin{eqnarray*}
e^{ip_M(\la)} &=& \frac{\la+i(M+\frac{1}{2})}{\la-i(M+\frac{1}{2})}\\
e^{i S_{MM'}(\la)} &=& \prod_{m=|M-M'|}^{M+M'}
\frac{\la+im}{\la-im}\cdot \frac{\la+i(m+1)}{\la-i(m+1)}.
\end{eqnarray*}
Taking the logarithm of (\ref{eq:bethe 2}) using the branch cut
\begin{equation*}
\frac{1}{i}\ln \frac{\la+ia}{\la-ia} = \pi -2\arctan \frac{\la}{a}
\end{equation*}
we obtain
\begin{equation}\label{eq:Q}
2N \arctan \frac{\la^M_j}{M+\frac{1}{2}} = 2\pi Q^M_j
 + \sum_{\stackrel{M',j'}{(M',j')\neq (M,j)}}
 \Phi_{MM'}(\la^M_j-\la^{M'}_{j'}),
\end{equation}
where
\begin{equation*}
\Phi_{MM'}(\la) = 2 \sum_{m=|M-M'|}^{M+M'}
 \left(\arctan \frac{\la}{m}+\arctan\frac{\la}{m+1}\right).
\end{equation*}
The first term on the right is absent for $m=0$.
Here $Q^M_j$ is an integer or
half-integer depending on the configuration.

In addition to the string hypothesis, we assume that the $Q^M_j$
classify the $\la$'s uniquely: $\la^M_j$ increases if $Q^M_j$
increases and in a given string no $Q^M_j$ coincide. As we will see
shortly with this assumption one obtains the correct number
of solutions to the Bethe equations (\ref{eq:bethe 1}).

Using $\arctan \pm \infty = \pm \frac{\pi}{2}$ we obtain from
(\ref{eq:Q}) putting $\la^M_j=\infty$
\begin{equation*}
Q^M_\infty = \frac{N}{2} -\bigl(2M+\frac{1}{2}\bigr)\bigl(m_{2M+1}-1\bigr)
- \sum_{M'\neq M}\bigl(2\min(M,M')+1\bigr) m_{2M'+1}.
\end{equation*}
Since there are $2M+1$ strings in a given string of length $2M+1$,
the maximal admissible $Q^M_{\mathrm{max}}$ is
\begin{equation*}
Q^M_{\mathrm{max}} = Q^M_\infty -(2M+1)
\end{equation*}
where we assume that if $Q^M_j$ is bigger than $Q^M_{\mathrm{max}}$
then at least one root in the string is infinite and hence all
are infinite which would imply $Q^M_j=Q^M_\infty$.

With the already mentioned assumption that each admissible set of
quantum number $Q^M_j$ corresponds uniquely to a solution of the
Bethe equations we may now count the number of Bethe vectors.
Since $\arctan$ is an odd function and by the assumption about the
monotonicity we have
\begin{equation*}
-Q^M_{\mathrm{max}}\le Q^M_1<\cdots <Q^M_{m_{2M+1}}\le Q^M_{\mathrm{max}}.
\end{equation*}
Hence defining $P_\ell$ as
\begin{equation*}
P_\ell=N-2\sum_{\ell'} \min(\ell,\ell')m_{\ell'}
\end{equation*}
so that
\begin{equation*}
P_\ell+m_\ell=2Q^M_{\max}+1 \qquad \mbox{with $\ell=2M+1$}.
\end{equation*}
With this the number of Bethe vectors with configuration
$\{m_\ell\}$ is given by
\begin{equation*}
Z(N,n|\{m_\ell\}) = \prod_{\ell\ge 1} 
 \bin{P_\ell+m_\ell}{m_\ell}
\end{equation*}
where $\bin{p+m}{m}=(p+m)!/p!m!$ is the binomial
coefficient. The total number of Bethe vectors is
\begin{equation}\label{eq:count}
Z(N,n) = \sum_{\stackrel{\{m_\ell\}}{\sum_\ell \ell m_\ell=n}}
 \prod_{\ell\ge 1} \bin{P_\ell+m_\ell}{m_\ell}.
\end{equation}

It should be emphasized that the derivation of (\ref{eq:count})
given here is not mathematically rigorous. Besides the various
assumptions that were made we also did not worry about possible
singularities of (\ref{eq:inter}). However, as we shall see
in the next section, (\ref{eq:count}) indeed yields the correct
number of Bethe vectors.

\section{Rigged configurations}
\label{sec:rig}

In the last section we parametrized the Bethe vectors by solutions
to the Bethe equations. As we have seen in section \ref{sec:model}
the state space is the tensor product of irreducible representations
of the underlying algebra, in our case the tensor product of
$\Complex^2$ with underlying algebra being $\su(2)$. The Bethe vectors
are the highest weight vectors in the irreducible components in this
tensor product.

In this section we will interpret (\ref{eq:count}) combinatorially in 
terms of rigged configurations. Since the Bethe vectors are
also the irreducible components of the underlying tensor product
which can be labeled by Young tableaux or crystal elements,
one may expect a bijection between the rigged configurations
and crystal elements. For the case $A_n$ such a bijection
is indeed known to exist \cite{KKR:1986,KR:1988,KSS:2002}.
For other types it was recently given in special cases in \cite{OSS:2002}.

To interpret (\ref{eq:count}) combinatorially let us view
the set $\{m_\ell\}$ as a partition $\nu$. A partition is a
set of numbers $\nu=(\nu_1,\nu_2,\ldots)$ such that $\nu_i\ge \nu_{i+1}$
and only finitely many $\nu_i$ are nonzero. The partition has
part $i$ if $\nu_k=i$ for some $k$. The size of partition $\nu$
is $|\nu|:=\nu_1+\nu_2+\cdots$. In the correspondence
between $\{m_\ell\}$ and $\nu$, $m_\ell$ specifies the number of parts
of size $\ell$ in $\nu$. For example, if $m_1=1$, $m_2=3$, $m_4=1$ and
all other $m_\ell=0$ then $\nu=(4,2,2,2,1)$.

It is well-known (see e.g. \cite{And:1976}) that 
$\bin{p+m}{m}$ is the number of partitions
in a box of size $p\times m$, meaning, that the partition cannot
have more than $m$ parts and no part exceeds $p$. Let
$\RC(N,n)$ be the set of all rigged configurations $(\nu,J)$ defined 
as follows. $\nu$ is a partition of size $|\nu|=n$ and $J$ is a
set of partition where $J_\ell$ is a partition in a box of
size $P_\ell\times m_\ell$. Then (\ref{eq:count}) can be rewritten as 
\begin{equation*}
Z(N,n) = \sum_{(\nu,J)\in \RC(N,n)} 1.
\end{equation*}

\begin{example}\label{ex:rig}
Let $N=5$ and $n=2$. Then the following is the set of
rigged configuration $\RC(5,2)$
\begin{eqnarray*}
&\begin{array}[t]{|c|c|c} \cline{1-2} \rig{} & \rig{1} & 1\\
 \cline{1-2} \end{array}
&\qquad
\begin{array}[t]{|c|c} \cline{1-1} \rig{1}&1\\ \cline{1-1} \rig{1}&\\ 
 \cline{1-1} \end{array}
\\
&\begin{array}[t]{|c|c|c} \cline{1-2} \rig{} & \rig{0} & 1\\
 \cline{1-2} \end{array}
&\qquad
\begin{array}[t]{|c|c} \cline{1-1} \rig{1}&1\\ \cline{1-1} \rig{0}&\\ 
 \cline{1-1} \end{array}
\\
&
&\qquad
\begin{array}[t]{|c|c} \cline{1-1} \rig{0}&1\\ \cline{1-1} \rig{0}&\\ 
 \cline{1-1}\end{array}
\end{eqnarray*}
The underlying partition on the left is (2) and on the right (1,1).
The partitions $J_\ell$ attached to part length $\ell$ is specified
by the numbers in each part. For example, the partition $J_1$ for the top
rigged configuration on the right is (1,1) whereas for the one in the
middle and bottom is $J_1=(1)$ and $J_1=\emptyset$, respectively.
The numbers to the right of part $\ell$ is $P_\ell$.
\end{example}

There exists a statistics on $\RC(N,n)$, called cocharge.
It is given by
\begin{equation*}
\cc(\nu,J) = \cc(\nu) + \sum_\ell |J_\ell|
\end{equation*}
where
\begin{equation*}
\cc(\nu) = \sum_{j,k} \min(j,k) m_j m_k.
\end{equation*}
For example, the cocharge for the rigged configurations in 
Example~\ref{ex:rig} from top to bottom, left to right is
3, 2, 6, 5, 4, respectively.

As mentioned before, rigged configurations are in bijection with
crystal elements. For our $\su(2)$ example these are all sequences
of 1's and 2's of length N such that the number of 2's never exceeds
the number of 1's reading the sequence from right to left. The
last condition is that of Yamanouchi words. The number
$n$ fixes the number of 2's in the sequence. Denote the set of all
such sequences by $\Path(N,n)$. For a path $p=p_N\cdots p_1\in \Path(N,n)$
define the energy as
\begin{equation}\label{eq:E}
E(p) = \sum_{j=1}^{N-1} (N-j) \chi(p_{j+1}>p_j)
\end{equation}
where $\chi(\mathrm{True})=1$ and $\chi(\mathrm{False})=0$.
The generating function of paths is given by
\begin{equation*}
X(N,n)=\sum_{p\in\Path(N,n)} q^{E(p)}.
\end{equation*}

\begin{example}\label{ex:path}
The set $\Path(5,2)$ is given by
\begin{equation*}
\Path(5,2)=\{22111,21211,12211,21121,12121\}.
\end{equation*}
The energies are 2, 4, 3, 5 and 6, respectively.
Hence $X(5,2)=q^2+q^3+q^4+q^5+q^6$. 
\end{example}

The bijection between $\Path(N,n)$ and $\RC(N,n)$ is defined recursively.
A path $p=p_N\cdots p_1\in \Path(N,n)$ is built up successively
from right to left. The empty path is mapped to the empty rigged
configuration. Assume that $p_{i-1}\cdots p_1$ corresponds to 
$(\nu^{i-1},J^{i-1})$. If $p_i=1$, $(\nu^i,J^i)=(\nu^{i-1},J^{i-1})$.
If $p_i=2$, then add a box to the largest singular string in
$(\nu^{i-1},J^{i-1})$ and make it singular again. A string is singular
if its label is equal to the vacancy number, in other words, if 
$J_\ell$ has a part of size $P_\ell$. In the final rigged configuration
$(\nu^N,J^N)$ take the complement of the partitions $J^N_\ell$
in the box $P^N_\ell\times m^N_\ell$. Let us call this map
$\Psi:\Path(N,n)\to\RC(N,n)$. We have the following 
theorem~\cite{KKR:1986,KR:1988,KSS:2002}.

\begin{theorem}
\label{theorem:bij su2}
The map $\Psi:\Path(N,n)\to \RC(N,n)$ is a bijection and
$E(p)=\cc(\Psi(p))$ for all $p\in\Path(N,n)$.
\end{theorem}

\begin{example}
Take $p=21121$. We get successively
\begin{equation*}
\begin{array}{ll}
p &\qquad (\nu,J)\\[2mm]
\emptyset &\qquad \emptyset\\[2mm]
1 &\qquad \emptyset\\[2mm]
21 &\qquad \begin{array}[c]{|c|c} \cline{1-1} \rig{0}&0\\ \cline{1-1}
 \end{array}\\[2mm]
121 &\qquad \begin{array}[c]{|c|c} \cline{1-1} \rig{0}&1\\ \cline{1-1}
 \end{array}\\[2mm]
1121 &\qquad \begin{array}[c]{|c|c} \cline{1-1} \rig{0}&2\\ \cline{1-1}
 \end{array}\\[2mm]
21121 &\qquad \begin{array}[c]{|c|c} \cline{1-1} \rig{1}&1\\ \cline{1-1}
 \rig{0}&\\ \cline{1-1} \end{array}
\end{array}
\end{equation*}
Hence $\Psi(21121)=\begin{array}[c]{|c|c} \cline{1-1} \rig{1}&1\\ \cline{1-1}
 \rig{0}&\\ \cline{1-1} \end{array}$. Similarly,
\begin{eqnarray*}
\Psi(22111) &=& \begin{array}[c]{|c|c|c} \cline{1-2} \rig{}&\rig{0}&1\\ 
 \cline{1-2}\end{array}\\
\Psi(21211) &=& \begin{array}[c]{|c|c} \cline{1-1} \rig{0}&1\\ \cline{1-1} 
 \rig{0}&\\ \cline{1-1} \end{array}\\
\Psi(12211) &=& \begin{array}[c]{|c|c|c} \cline{1-2} \rig{}&\rig{1}&1\\ 
 \cline{1-2}\end{array}\\
\Psi(12121) &=& \begin{array}[c]{|c|c} \cline{1-1} \rig{1}&1\\ \cline{1-1}
 \rig{1}&\\ \cline{1-1} \end{array}
\end{eqnarray*}
Comparing with examples \ref{ex:rig} and \ref{ex:path}, the statistics
match.
\end{example}

It follows immediately from Theorem \ref{theorem:bij su2} that
\begin{equation*}
X(N,n) = \sum_{(\nu,J)\in \RC(N,n)} q^{\cc(\nu,J)}.
\end{equation*}
The $q$-binomial coefficient
\begin{equation*}
\qbin{p+m}{m} = \frac{(q)_{p+m}}{(q)_p(q)_m},
\end{equation*}
where $(q)_m=\prod_{i=1}^m (1-q^i)$, is the generating function
of partitions in a box of size $p\times m$ \cite{And:1976}.
Hence, defining $\Conf(N,n)$ to be the set of all partitions
$\nu$ of $n$ such that $P_\ell\ge 0$ for all $\ell$ the following
corollary holds. The right-hand side is called fermionic formula.
\begin{corollary}\label{cor:fermi}
\begin{equation*}
X(N,n) = \sum_{\nu\in \Conf(N,n)} q^{\cc(\nu)}
 \prod_\ell \qbin{P_\ell+m_\ell}{m_\ell}.
\end{equation*}
\end{corollary} 

\section{Generalizations}
\label{sec:general}

So far we have only considered the spin 1/2 XXX model and its
counting. This model is based on the fundamental representation
of $\mathfrak{su}(2)$. It turns out that the $q$-counting of
Corollary \ref{cor:fermi} is associated with the Kac--Moody
Lie algebra $A_1^{(1)}$. In the remainder of this note we will
indicate how to generalize the $q$-counting that arises from the
Bethe Ansatz.

The set of paths $\Path(N,n)$, which is the set of Yamanouchi words
in the letters 1 and 2 of length $N$ with $n$ twos, will be generalized
to the set of highest weight elements in a tensor product of
crystals of a given weight; the Yamanouchi condition is replaced by
the highest weight condition and the condition on the number of
twos becomes the requirement on the weight.
Crystal bases were first introduced by Kashiwara \cite{K:1990}
in connection with quantized universal enveloping algebras.
The quantized universal enveloping algebra $U_q(\gggg)$ associated
with a symmetrizable Kac--Moody Lie algebra $\gggg$ was discovered
independently by Drinfeld \cite{D:1985} and Jimbo \cite{J:1985} in their
study of two dimensional solvable lattice models in statistical
mechanics. The parameter $q$ corresponds to the temperature of the
underlying model. Kashiwara \cite{K:1990} showed that at zero
temperature or $q=0$ the representations of $U_q(\gggg)$ have
bases, which he coined crystal bases, with a beautiful
combinatorial structure and favorable properties such as
uniqueness and stability under tensor products.

In the generalization from $\mathfrak{su}(2)$ to other types,
rigged configurations become sequences of partitions with
riggings. The number of partitions depends on the rank of the
underlying algebra.

The generalization of the bijection from paths to rigged
configurations to type $A_n^{(1)}$ is given in \cite{KR:1988,KSS:2002}
and to other nonexceptional types in \cite{OSS:2002} in special cases.
It was shown in \cite{OSS:2001,OSS:2002a} that all crystals can be
realized as crystals of simply-laced type $A,D,E$. Hence the
bijections for these types can be viewed as fundamental.

In the next section we will introduce crystal bases.
The bijection algorithm for type $A_n^{(1)}$ and $D_n^{(1)}$
is presented in section \ref{sec:bij}.

\section{Crystals}
\label{sec:crystals}

\subsection{Axiomatic definition of crystals}
Let $\gggg$ be an affine Lie algebra and $I$ the index set of its
Dynkin diagram. Let $\alpha_i,h_i,\La_i$ ($i\in I$) be the simple roots, simple
coroots, and fundamental weights for $\gggg$. Let
$\delta=\sum_{i\in I}a_i\alpha_i$ denote the standard null root
and $c=\sum_{i\in I}a_i^\vee h_i$ the canonical central element,
where $a_i,a_i^\vee$ are the positive integers given in
\cite{Kac}. Let $P=\bigoplus_{i\in I}\Z\La_i\oplus\Z\delta$ be the
weight lattice and $P^+=\sum_{i\in I}\Zn\La_i\oplus\Z\delta$ the
dominant weights.

A crystal $B$ is a set $B=\sqcup_{\la\in P}B_\la$ ($\wt\, b=\la$
if $b\in B_\la$) with the maps
\begin{eqnarray*}
&&e_i: B_\la\longrightarrow B_{\la+\alpha_i}\sqcup\{0\},\quad
 f_i: B_\la\longrightarrow B_{\la-\alpha_i}\sqcup\{0\},\\
&&\veps_i : B\longrightarrow \Z\sqcup\{-\infty\},\quad
\vphi_i : B\longrightarrow \Z\sqcup\{-\infty\}
\end{eqnarray*}
for all $i\in I$ such that
\begin{eqnarray*}
&&\mbox{for $b\in B_\la$, $\vphi_i(b)=\langle h_i,\la\rangle+\veps_i(b)$},\\
&&\mbox{for $b\in B$, we have}\\
&&\hspace{1.5cm}
\begin{array}{rcl}
\veps_i(b)&=&\veps_i(e_ib)+1\mbox{ if }e_i b\neq0,\\
&=&\veps_i(f_i b)-1\mbox{ if }f_i b\neq0,\\
\vphi_i(b)&=&\vphi_i(e_i b)-1\mbox{ if } e_i b\neq0,\\
&=&\vphi_i(f_i b)+1\mbox{ if }f_i b\neq0,
\end{array} \nonumber\\
&&\mbox{for $b,b'\in B$, $e_i b'=b$ if and only if $b'=f_i b$},\\
&&\mbox{for $b\in B$, $\veps_i(b)=\vphi_i(b)=-\infty$ implies
$e_i b=f_i b=0$}.
\end{eqnarray*}
A crystal $B$ can be regarded as a colored oriented graph by defining
\begin{equation*}
b\stackrel{i}{\longrightarrow}b'\quad\Longleftrightarrow\quad f_ib=b'.
\end{equation*}
If we want to emphasize $I$, $B$ is called an $I$-crystal.

If $B_1$ and $B_2$ are crystals, then for $b_1\otimes b_2\in B_1\otimes B_2$
the action of $e_i$ is defined as
\begin{equation*}
e_i(b_1\otimes b_2)=\left\{ \begin{array}{ll}
e_ib_1 \otimes b_2 &\mbox{if $\varepsilon_i(b_1)>\varphi_i(b_2)$,}\\
b_1\otimes e_i b_2 &\mbox{else,}
\end{array} \right.
\end{equation*}
where $\varepsilon_i(b)=\max\{k\mid e_i^k b \;\mbox{is defined}\}$ and
$\varphi_i(b)=\max\{k\mid f_i^k b\;\mbox{is defined}\}$. This is the opposite
of the notation used by Kashiwara \cite{K:1990}.

An element $b\in B$ is classically highest weight if $e_i b=0$ for
all $i=1,2,\ldots,n$. For $B=B_L\otimes \cdots \otimes B_1$ and
$\La\in P^+$, the set of paths is defined as follows
\begin{equation*}
\Path(B,\La)=\{b\in B \mid \mbox{$e_ib=0$ for all $i=1,2,\ldots,n$,
$\wt b =\La$}\}.
\end{equation*}

In the following we will discuss the crystals of type $A_n^{(1)}$ and 
$D_n^{(1)}$ more explicitly.

\subsection{Dynkin data of type $A_n$ and $D_n$}
Let $\epsilon_i$ be the $i$-th standard unit vector in $\Z^n$.
Then for type $A_{n-1}$, the simple roots are
\begin{equation*}
\alpha_i=\epsilon_i-\epsilon_{i+1} \qquad \mbox{for $1\le i<n$}
\end{equation*}
and the fundamental weights are
\begin{equation*}
\La_i=\epsilon_1+\cdots+\epsilon_i \qquad \mbox{for $1\le i< n$.}
\end{equation*}
For type $D_n$, the simple roots are
\begin{eqnarray*}
\alpha_i&=&\epsilon_i-\epsilon_{i+1} \qquad \mbox{for $1\le i<n$}\\
\alpha_n&=&\epsilon_{n-1}+\epsilon_n
\end{eqnarray*}
and the fundamental weights are
\begin{eqnarray*}
\La_i&=&\epsilon_1+\cdots+\epsilon_i \qquad\qquad\mbox{for $1\le i\le n-2$}\\
\La_{n-1}&=&(\epsilon_1+\cdots+\epsilon_{n-1}-\epsilon_n)/2 \\
\La_n&=&(\epsilon_1+\cdots+\epsilon_{n-1}+\epsilon_n)/2.
\end{eqnarray*}

\subsection{Affine crystals of type $A_n^{(1)}$ and $D_n^{(1)}$}
In \cite{HKOTY:1999} it is conjectured that there is a family of
finite-dimensional irreducible $U'_q(\gggg)$-modules
$\{W^{(a)}_i\mid a\in J,i\in \Zn\}$ which, unlike most
finite-dimensional $U'_q(\gggg)$-modules, have crystal bases
$B^{a,i}$. Here $U'_q(\gggg)$ is the quantum universal enveloping algebra
of the derived subalgebra of $\gggg$, obtained by omitting the degree
operator, and $J=I\backslash\{0\}$. 

\begin{figure}
\begin{tabular}{|c|l|}
\hline
%
$A_n^{(1)}$ & \raisebox{-0.7cm}{\scalebox{0.7}{
\begin{picture}(250,62)(-10,-12)
\BText(0,0){1} \LongArrow(10,0)(40,0) \BText(50,0){2}
\LongArrow(60,0)(90,0) \BText(100,0){3} \LongArrow(110,0)(140,0)
\Text(160,0)[]{$\cdots$} \LongArrow(175,0)(205,0)
\BText(220,0){n+1} \LongArrowArc(110,-181)(216,62,118)
\PText(25,2)(0)[b]{1} \PText(75,2)(0)[b]{2} \PText(125,2)(0)[b]{3}
\PText(190,2)(0)[b]{n} \PText(110,38)(0)[b]{0}
\end{picture}
}}
\\ \hline
%
$D_n^{(1)}$ & \raisebox{-1.3cm}{\scalebox{0.7}{
\begin{picture}(365,100)(-10,-50)
\BText(0,0){1} \LongArrow(10,0)(30,0) \BText(40,0){2}
\LongArrow(50,0)(70,0) \Text(85,0)[]{$\cdots$}
\LongArrow(95,0)(115,0) \BText(130,0){n-1}
\LongArrow(143,2)(160,14) \LongArrow(143,-2)(160,-14)
\BText(170,15){n} \BBoxc(170,-15)(13,13)
\Text(170,-15)[]{\footnotesize$\overline{\mbox{n}}$}
\LongArrow(180,14)(197,2) \LongArrow(180,-14)(197,-2)
\BBoxc(215,0)(25,13)
\Text(215,0)[]{\footnotesize$\overline{\mbox{n-1}}$}
\LongArrow(230,0)(250,0) \Text(265,0)[]{$\cdots$}
\LongArrow(275,0)(295,0) \BBoxc(305,0)(13,13)
\Text(305,0)[]{\footnotesize$\overline{\mbox{2}}$}
\LongArrow(315,0)(335,0) \BBoxc(345,0)(13,13)
\Text(345,0)[]{\footnotesize$\overline{\mbox{1}}$}
\LongArrowArc(192.5,-367)(402,69,111)
\LongArrowArcn(152.5,367)(402,-69,-111) \PText(20,2)(0)[b]{1}
\PText(60,2)(0)[b]{2} \PText(105,2)(0)[b]{n-2}
\PText(152,13)(0)[br]{n-1} \PText(152,-9)(0)[tr]{n}
\PText(188,13)(0)[bl]{n} \PText(188,-9)(0)[tl]{n-1}
\PText(240,2)(0)[b]{n-2} \PText(285,2)(0)[b]{2}
\PText(325,2)(0)[b]{1} \PText(192.5,38)(0)[b]{0}
\PText(152.5,-35)(0)[t]{0}
\end{picture}
}}
\\ \hline
\end{tabular}\vspace{4mm}
\caption{\label{fig:crystals}Crystals $B^{1,1}$}
\end{figure}
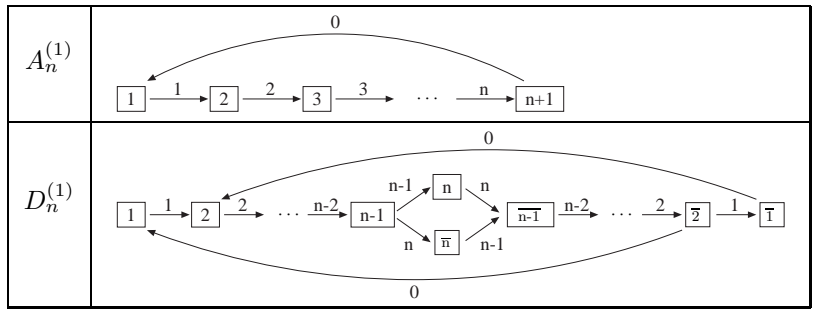

Here we will restrict our attention to the simplest affine crystals
$B^{1,1}$ of type $A_n^{(1)}$ and $D_n^{(1)}$. As a set $B^{1,1}$
is $\{1<2<\cdots<n+1\}$ for type $A_n^{(1)}$ and 
$\{1<2<\cdots<n-1<\begin{array}{c} n\\ \overline{n}\end{array}
 <\overline{n-1}<\cdots<\overline{1}\}$ for type $D_n^{(1)}$. 
The crystal graphs are given in Figure \ref{fig:crystals}.

\subsection{One-dimensional sums}
The energy function (\ref{eq:E}) can be generalized to the crystal
setting. In the case $B=(B^{1,1})^{\otimes L}$ it takes a simple form. 
There is \cite{KMN:1992} a unique (up to global additive constant) function 
$H:B^{1,1}\otimes B^{1,1}\rightarrow\Z$ called the local energy function, 
such that
\begin{equation*}
  H(e_i(b\otimes b')) = H(b\otimes b')+
  \left\{\begin{array}{ll}
  -1 & \mbox{if $i=0$ and $e_0(b\otimes b')=b\otimes e_0 b'$}
  \\
  1 & \mbox{if $i=0$ and $e_0(b\otimes b')=e_0 b\otimes b'$}
  \\
  0 & \mbox{otherwise.}
  \end{array}\right.
\end{equation*}
We normalize $H$ by the condition $H(1 \otimes 1) = 0$.
\begin{example}
Let $b\otimes b'\in B^{1,1}\otimes B^{1,1}$. Explicitly,
the local energy function is given as follows.
For type $A_n^{(1)}$, $H(b\otimes b')=-\chi(b>b')$.
For type $D_n^{(1)}$, $H(b\otimes b')=0$ if $b\le b'$,
$H(b\otimes b')=-1$ if $b\otimes b'=n\otimes \overline{n},
\overline{n}\otimes n$ or $b>b'$ where $b\otimes b'\neq 
\overline{1}\otimes 1$, and $H(\overline{1}\otimes 1)=-2$.
\end{example}

For $b_L\otimes\cdots\otimes b_1\in B=(B^{1,1})^{\otimes L}$
\begin{equation*}
  E(b_L\otimes\dots \otimes b_1) =
  \sum_{j=1}^{L-1} (L-j) \,\,H(b_{j+1}\otimes b_j).
\end{equation*}
Define the one-dimensional sum $X(B,\la;q)\in\Z[q,q^{-1}]$ by
\begin{equation*}
  X(B,\la;q) = \sum_{b\in \Path(B,\la)} q^{E(b)}.
\end{equation*}
\begin{example}
In the crystal language the set of paths of Example \ref{ex:path}
corresponds to $B=(B^{1,1})^{\otimes 5}$ of type $A_1^{(1)}$ of
weight $\la=\La_1+2\La_2$.
\end{example}

\section{Fermionic formula and rigged configurations}
\label{sec:rc}

Fermionic formulas associated to a Kac-Moody algebra $\gggg$
were conjectured in \cite{HKOTT:2001,HKOTY:1999}. 
We review the fermionic formulas for type $A_n^{(1)}$ and $D_n^{(1)}$.

Let $L_i^{(a)}$ with $a\in J$ and $i\in \Zn$ denote the number of tensor 
factors $B^{a,i}$ in $B$ and let $\la$ be a dominant integral weight. 
Say that $\nt=(m_i^{(a)})$ is a $(B,\la)$-configuration if
\begin{equation}\label{eq:constraint}
\sum_{\stackrel{a\in J}{i\in\Zn}} i\,
m_i^{(a)} \alpha_a = \sum_{\stackrel{a\in J}{i\in\Zn}} 
i \,L_i^{(a)} \La_a - \la.
\end{equation}
The configuration $\nt$ is admissible if all vacancy numbers are nonnegative
\begin{equation*}
  p_i^{(a)} \ge 0\qquad\mbox{for all $a\in J$ and $i\in\Zn$,}
\end{equation*}
where
\begin{equation}\label{eq:p}
p_i^{(a)} = \sum_{k\in\Zn} \left( L_k^{(a)} \min(i,k) -
\sum_{b\in J} (\alpha_a|\alpha_b) \min(i,k)\, m_k^{(b)}\right).
\end{equation}
Write $C(B,\la)$ for the set of admissible $(B,\la)$-configurations. Define
\begin{equation*}
cc(\nt) = \frac{1}{2} \sum_{a,b\in J} \sum_{j,k\in\Zn} 
(\alpha_a|\alpha_b) \min(j,k) m_j^{(a)} m_k^{(b)}.
\end{equation*}

The fermionic formula is defined by
\begin{equation}\label{eq:fermi}
M(B,\la;q) = \sum_{\nt\in C(B,\la)} q^{cc(\nt)}
\prod_{a\in J} \prod_{i\in\Zn}
\qbin{p_i^{(a)}+m_i^{(a)}}{m_i^{(a)}}.
\end{equation}

The $X=M$ conjecture of \cite{HKOTT:2001,HKOTY:1999} states that
\begin{equation*}
  X(B,\la;q^{-1})=M(B,\la;q).
\end{equation*}

The fermionic formula $M(B,\la)$ can be interpreted using
rigged configurations.
Denote by $(\nt,\Jt)$ a pair where $\nt=(m_i^{(a)})$ is a matrix and
$\Jt=(J^{(a,i)})$ is a matrix of partitions with $a\in J$ and $i\in\Zn$.
Then a rigged configuration is a pair $(\nt,\Jt)$ such that $\nt\in C(B,\la)$
and the partition $J^{(a,i)}$ is contained in a $m_i^{(a)}\times
p_i^{(a)}$ rectangle for all $a,i$. The set of rigged
$(B,\la)$-configurations for
fixed $\la$ and $B$ is denoted by $\RC(B,\la)$. Then
(\ref{eq:fermi}) is equivalent to
\begin{equation*}
M(B,\la)=\sum_{(\nt,\Jt)\in\RC(B,\la)} q^{\cc(\nt,\Jt)}
\end{equation*}
where $\cc(\nt,\Jt)=\cc(\nt)+|\Jt|$
and $|\Jt|=\sum_{(a,i)} |J^{(a,i)}|$.
To emphasize the dependence on $\nt$ we also write
$m_i^{(a)}(\nt)$ and $P_i^{(a)}(\nt)$ for $m_i^{(a)}$ and $p_i^{(a)}$,
respectively.

\section{Bijection between rigged configurations and paths}
\label{sec:bij}

In this section we give the description of the bijection 
$\Phi:\RC(B,\la)\to\Path(B,\la)$ for types $A_n^{(1)}$ and $D_n^{(1)}$
when $B=(B^{1,1})^{\otimes L}$. 

Let $(\nt,\Jt)\in\RC(B,\la)$. We shall define a map
$\rk:\RC(B,\la)\to B^{1,1}$ which associates to $(\nt,\Jt)$ an element
of $B^{1,1}$ called its rank.
Denote by $\RC_b(B,\la)$ the elements of $\RC(B,\la)$ of rank
$b$. We shall define a bijection
$\delta:\RC_b(B,\la)\to\RC(\Bt,\la-\wt(b))$ where
$\Bt=(B^{1,1})^{\otimes (L-1)}$. The disjoint union
of these bijections then defines a bijection
$\delta:\RC(B,\la)\to\bigcup_{b\in B^{1,1}} \RC(\Bt,\la-\wt(b))$.

The bijection $\Phi$ is defined recursively as follows. For $b\in
B^{1,1}$ let $\Path_b(B,\la)$ be the set of paths in $B$
that have $b$ as leftmost tensor factor.
For $L=0$ the bijection $\Phi$ sends the empty rigged
configuration (the only element of the set $\RC(B,\la)$) to the
empty path (the only element of $\Path(B,\la)$). Otherwise
assume that $\Phi$ has been defined for $\Bt$ and
define it for $B$ by the commutative diagram
\begin{equation*}
\begin{CD}
\RC_b(B,\la) @>{\Phi}>> \Path_b(B,\la) \\
@V{\delta}VV @VVV \\
\RC(\Bt,\la-\wt(b)) @>{\Phi}>> \Path(\Bt,\la-\wt(b))
\end{CD}
\end{equation*}
where the right hand vertical map removes the leftmost tensor
factor $b$. In short,
\begin{equation*}
  \Phi(\nt,\Jt)=\rk(\nt,\Jt)\otimes \Phi(\delta(\nt,\Jt)).
\end{equation*}

We also require the bijection
$\Phit:\RC(B,\la)\to\Path(B,\la)$ given by $\Phit=\Phi\circ
\comp$ where $\comp:\RC(B,\la)\to\RC(B,\la)$ with
$\comp(\nt,\Jt)=(\nt,\Jtt)$ is the function which complements the
riggings, meaning that $\Jtt$ is obtained from $\Jt$ by
complementing all partitions $J^{(a,i)}$ in the $m_i^{(a)}(\nt)\times
P_i^{(a)}(\nt)$ rectangle.

\begin{remark}
The bijection $\Psi$ of section \ref{sec:rig} is the inverse of
$\Phit$ for type $A_1^{(1)}$.
\end{remark}

\begin{theorem}\label{thm:bij}
$\Phi:\RC(B,\la)\to\Path(B,\la)$ is a bijection such that
\begin{equation*}
cc(\nt,\Jt)=-E(\Phit(\nt,\Jt))\qquad\mbox{for all
$(\nt,\Jt)\in\RC(B,\la)$.}
\end{equation*}
\end{theorem}
For type $A_n^{(1)}$ a generalization of this theorem for
$B=B^{a_L,i_L}\otimes \cdots\otimes B^{a_1,i_1}$
was proven in \cite{KSS:2002}.
For other types Theorem \ref{thm:bij} is proved in \cite{OSS:2002}.

To describe the bijection explicitly for types $A_n^{(1)}$ and $D_n^{(1)}$,
the following notation is needed.
The matrix $\nt=(m_i^{(a)})$ can be viewed as a sequence of partitions
$\nt=(\nu^{(1)},\nu^{(2)},\ldots,\nu^{(n)})$ where
$m_i^{(a)}$ is the number of parts of size $i$ in the partition
$\nu^{(a)}$. Denote by $Q_i(\rho)$ the number of boxes in the first
$i$ columns of the partition $\rho$. Finally the partition
$J^{(a,i)}$ is called singular if it has a part of size $p_i^{(a)}$.

\subsection{Bijection for type $A_n^{(1)}$}

Using the Dynkin data for type $A_n$ the vacancy numbers (\ref{eq:p})
and the constraints (\ref{eq:constraint}) can be rewritten in the following
explicit way
\begin{equation*}
P_i^{(a)}(\nt)=Q_i(\nu^{(a-1)})-2Q_i(\nu^{(a)})+Q_i(\nu^{(a+1)})
 +L\delta_{a,1} \qquad\mbox{for $1\le a\le n$}
\end{equation*}
and
\begin{equation*}
|\nu^{(a)}|=L-\sum_{b=1}^a \la_b \qquad\mbox{for $1\le a\le n$.}
\end{equation*}

The algorithm $\delta$ is given as follows.
Set $\ell^{(0)}=0$ and repeat the following process for
$a=1,2,\ldots,n$ or until stopped. Find the minimal index $i\ge
\ell^{(a-1)}$ such that $J^{(a,i)}$ is singular. If no such $i$
exists, set $b=a$ and stop. Otherwise set $\ell^{(a)}=i$ and
continue with $a+1$. If the process did not stop, set $b=n+1$.
Set all undefined $\ell^{(a)}$ to $\infty$.

The new rigged configuration is defined by
\begin{equation*}
 m_i^{(a)}(\ntt)=m_i^{(a)}(\nt)+\left\{ \begin{array}{ll}
 1 & \mbox{if $i=\ell^{(a)}-1$}\\
 -1 & \mbox{if $i=\ell^{(a)}$}\\
 0 & \mbox{otherwise.} \end{array}\right.
\end{equation*}
The partition $\tilde{J}^{(a,i)}$ is obtained from $J^{(a,i)}$ by removing
a part of size $P_i^{(a)}(\nt)$ for $i=\ell^{(a)}$, adding a part of size 
$P_i^{(a)}(\ntt)$ for $i=\ell^{(a)}-1$, and leaving it unchanged otherwise.

\begin{example}
Take $B=(B^{1,1})^{\otimes 7}$, $\la=\La_3+\La_4$ and 
$(\nt,\Jt)\in \RC(B,\la)$ as
\begin{equation*}
\Bigl(
\raisebox{0.5cm}{
\begin{array}[t]{|c|c|l} \cline{1-2} \rig{}&\rig{0}&0\\ \cline{1-2} 
 \rig{}&\rig{0}& \\
 \cline{1-2} \rig{3}&\multicolumn{2}{l}{3} \\ \cline{1-1} \end{array}},
\raisebox{0.5cm}{
\begin{array}[t]{|c|c|l} \cline{1-2} \rig{}&\rig{0}&0\\ \cline{1-2} 
 \rig{0}&\multicolumn{2}{l}{0}\\
 \cline{1-1} \end{array}},
\raisebox{0.5cm}{
\begin{array}[t]{|c|l} \cline{1-1} \rig{0}&0\\ \cline{1-1} \end{array}}
\Bigr).
\end{equation*}
The algorithm for $\Phi$ on $\comp(\nt,\Jt)$ yields
\begin{equation*}
\begin{array}{lll|c}
(\nt,\Jt)^{(1)} & (\nt,\Jt)^{(2)} & (\nt,\Jt)^{(3)} & \rk\\ \hline
\begin{array}[t]{|c|c|l} \cline{1-2} \rig{}&\rig{0}&0\\ \cline{1-2} 
 \rig{}&\rig{0}& \\ \cline{1-2} \rig{0}&\multicolumn{2}{l}{3} \\ \cline{1-1} 
 \end{array}
&
\begin{array}[t]{|c|c|l} \cline{1-2} \rig{}&\rig{0}&0\\ \cline{1-2} 
 \rig{0}&\multicolumn{2}{l}{0}\\
 \cline{1-1} \end{array}
&
\begin{array}[t]{|c|l} \cline{1-1} \rig{0}&0\\ \cline{1-1} \end{array}
&\\
\begin{array}[t]{|c|c|l} \cline{1-2} \rig{}&\rig{0}&0\\ \cline{1-2} 
 \rig{2}&\multicolumn{2}{l}{2}\\
 \cline{1-1} \rig{0}&\multicolumn{2}{l}{} \\ \cline{1-1} \end{array}
&
\begin{array}[t]{|c|l} \cline{1-1} \rig{0}&0\\ \cline{1-1} \rig{0}&\\ 
 \cline{1-1}\end{array}
&
\begin{array}[t]{|c|l} \cline{1-1} \rig{0}&0\\ \cline{1-1} \end{array}
&3\\
\begin{array}[t]{|c|c|l} \cline{1-2} \rig{}&\rig{0}&0\\ \cline{1-2} 
 \rig{0}&\multicolumn{2}{l}{2}\\ \cline{1-1} \end{array}
&
\begin{array}[t]{|c|l} \cline{1-1} \rig{0}&0\\ \cline{1-1} \end{array}
&\es
&4\\
\begin{array}[t]{|c|l} \cline{1-1} \rig{1}&1\\ \cline{1-1} \rig{0}&\\ 
 \cline{1-1}\end{array}
&
\begin{array}[t]{|c|l} \cline{1-1} \rig{0}&0\\ \cline{1-1} \end{array}
&\es
&2\\
\begin{array}[t]{|c|l} \cline{1-1} \rig{0}&1\\ \cline{1-1} \end{array}
&\es
&\es
&3\\
\begin{array}[t]{|c|l} \cline{1-1} \rig{0}&0\\ \cline{1-1} \end{array}
&\es
&\es
&1\\
\es&\es&\es&2\\
\es&\es&\es&1
\end{array}
\end{equation*}
Hence $\Phit(\nt,\Jt)=b=3\otimes 4\otimes 2\otimes 3\otimes 1\otimes 2
\otimes 1$ and $E(b)=\cc(\nt,\Jt)=12$.
\end{example}

\subsection{Bijection for type $D_n^{(1)}$}

Using the Dynkin data for type $D_n$ the vacancy numbers (\ref{eq:p})
and the constraints (\ref{eq:constraint}) can be rewritten in the following
explicit way
\begin{eqnarray*}
P_i^{(a)}(\nt)&=&Q_i(\nu^{(a-1)})-2Q_i(\nu^{(a)})+Q_i(\nu^{(a+1)})
 +L\delta_{a,1} \qquad\mbox{for $1\le a<n-2$}\\
P_i^{(n-2)}(\nt)&=&Q_i(\nu^{(n-3)})-2Q_i(\nu^{(n-2)})+Q_i(\nu^{(n-1)})
 +Q_i(\nu^{(n)})\\
P_i^{(n-1)}(\nt)&=&Q_i(\nu^{(n-2)})-2Q_i(\nu^{(n-1)})\\
P_i^{(n)}(\nt)&=&Q_i(\nu^{(n-2)})-2Q_i(\nu^{(n)})
\end{eqnarray*}
and
\begin{eqnarray*}
|\nu^{(a)}|&=&L-\sum_{b=1}^a \la_b \qquad\mbox{for $1\le a\le n-2$}\\
|\nu^{(n-1)}|&=&\frac{1}{2}(L-\sum_{b=1}^{n-1} \la_b+\la_n)\\
|\nu^{(n)}|&=&\frac{1}{2}(L-\sum_{b=1}^n \la_b).
\end{eqnarray*}

The algorithm $\delta$ is given as follows.
Set $\ell^{(0)}=0$ and repeat the following process for
$a=1,2,\ldots,n-2$ or until stopped. Find the minimal index $i\ge
\ell^{(a-1)}$ such that $J^{(a,i)}$ is singular. If no such $i$
exists, set $b=a$ and stop. Otherwise set $\ell^{(a)}=i$ and
continue with $a+1$.

If the process has not stopped at $a=n-2$ continue as follows.
Find the minimal indices $i,j\ge \ell^{(n-2)}$ such that
$J^{(n-1,i)}$ and $J^{(n,j)}$ are singular. If neither $i$ nor
$j$ exist, set $b=n-1$ and stop.
If $i$ exists, but not $j$, set $\ell^{(n-1)}=i$, $b=n$ and stop.
If $j$ exists, but not $i$, set $\ell^{(n)}=j$, $b=\overline{n}$
and stop. If both $i$ and $j$ exist, set $\ell^{(n-1)}=i$, $\ell^{(n)}=j$
and continue with $a=n-2$.

Now continue for $a=n-2,n-3,\ldots,1$ or until stopped.
Find the minimal index $i\ge \lb^{(a+1)}$ where $\lb^{(n-1)}
=\max(\ell^{(n-1)},\ell^{(n)})$ such that $J^{(a,i)}$ is singular
(if $i=\ell^{(a)}$ then there need to be two parts of size
$P_i^{(a)}(\nt)$ in $J^{(a,i)}$).
If no such $i$ exists, set $b=\overline{a+1}$ and stop.
If the process did not stop, set $b=\overline{1}$.

Set all yet undefined $\ell^{(a)}$ and $\lb^{(a)}$ to $\infty$.

The new rigged configuration is defined by
\begin{equation*}
 m_i^{(a)}(\ntt)=m_i^{(a)}(\nt)+\left\{ \begin{array}{ll}
 1 & \mbox{if $i=\ell^{(a)}-1$}\\
 -1 & \mbox{if $i=\ell^{(a)}$}\\
 1 & \mbox{if $i=\lb^{(a)}-1$ and $1\le a\le n-2$}\\
 -1 & \mbox{if $i=\lb^{(a)}$ and $1\le a \le n-2$}\\
 0 & \mbox{otherwise} \end{array}\right.
\end{equation*}
The partition $\tilde{J}^{(a,i)}$ is obtained from $J^{(a,i)}$ by removing
a part of size $P_i^{(a)}(\nt)$ for $i=\ell^{(a)}$ and $i=\lb^{(a)}$,
adding a part of size $P_i^{(a)}(\ntt)$ for $i=\ell^{(a)}-1$ and
$i=\lb^{(a)}-1$, and leaving it unchanged otherwise.

\begin{example}
Take $B=(B^{1,1})^{\otimes 6}$, $\la=2\La_3$ and $(\nt,\Jt)\in \RC(B,\la)$
as
\begin{equation*}
\Bigl(
\raisebox{0.5cm}{
\begin{array}[t]{|c|c|l} \cline{1-2} \rig{}&\rig{0}&0\\ \cline{1-2} 
 \rig{}&\rig{0}& \\
 \cline{1-2} \rig{0}&\multicolumn{2}{l}{2} \\ \cline{1-1} \end{array}},
\raisebox{0.5cm}{
\begin{array}[t]{|c|c|l} \cline{1-2} \rig{}&\rig{0}&0\\ \cline{1-2} 
 \rig{}&\rig{0}&\\ \cline{1-2} \end{array}},
\raisebox{0.5cm}{
\begin{array}[t]{|c|l} \cline{1-1} \rig{0}&0\\ \cline{1-1} \end{array}},
\raisebox{0.5cm}{
\begin{array}[t]{|c|c|l} \cline{1-2} \rig{}&\rig{0}&0\\ \cline{1-2} 
 \end{array}}
\Bigr).
\end{equation*}
Then the algorithm for $\Phi$ on $\comp(\nt,\Jt)$ gives the following 
intermediate steps
\begin{equation*}
\begin{array}{llll|c}
(\nt,\Jt)^{(1)} & (\nt,\Jt)^{(2)} & (\nt,\Jt)^{(3)} & (\nt,\Jt)^{(4)}
& \rk\\ \hline
\begin{array}[t]{|c|c|l} \cline{1-2} \rig{}&\rig{0}&0\\ \cline{1-2} 
 \rig{}&\rig{0}& \\
 \cline{1-2} \rig{2}&\multicolumn{2}{l}{2} \\ \cline{1-1} \end{array}
&
\begin{array}[t]{|c|c|l} \cline{1-2} \rig{}&\rig{0}&0\\ \cline{1-2} 
 \rig{}&\rig{0}&\\ \cline{1-2} \end{array}
&
\begin{array}[t]{|c|l} \cline{1-1} \rig{0}&0\\ \cline{1-1} \end{array}
&
\begin{array}[t]{|c|c|l} \cline{1-2} \rig{}&\rig{0}&0\\ \cline{1-2} \end{array}
&\\
\begin{array}[t]{|c|c|l} \cline{1-2} \rig{}&\rig{0}&0\\ \cline{1-2} 
 \rig{}&\rig{0}&\\ \cline{1-2} \end{array}
&
\begin{array}[t]{|c|c|l} \cline{1-2} \rig{}&\rig{0}&0\\ \cline{1-2} 
 \rig{0}& \multicolumn{2}{l}{0}\\ \cline{1-1} \end{array}
&
\begin{array}[t]{|c|l} \cline{1-1} \rig{0}&0\\ \cline{1-1} \end{array}
&
\begin{array}[t]{|c|l} \cline{1-1} \rig{0}&0\\ \cline{1-1} \end{array}
& \overline{4}\\
\begin{array}[t]{|c|c|l} \cline{1-2} \rig{}&\rig{0}&0\\ \cline{1-2} 
 \rig{2}& \multicolumn{2}{l}{2}\\ \cline{1-1} \end{array}
&
\begin{array}[t]{|c|l} \cline{1-1} \rig{0}&0\\ \cline{1-1} \rig{0}&\\ 
 \cline{1-1} \end{array}
&
\begin{array}[t]{|c|l} \cline{1-1} \rig{0}&0\\ \cline{1-1} \end{array}
&
\begin{array}[t]{|c|l} \cline{1-1} \rig{0}&0\\ \cline{1-1} \end{array}
& 3\\
\begin{array}[t]{|c|l} \cline{1-1} \rig{1}&1\\ \cline{1-1} \end{array}
&\es&\es&\es& \overline{1}\\
\es&\es&\es&\es& 2\\
\es&\es&\es&\es& 1\\
\es&\es&\es&\es& 1
\end{array}
\end{equation*}
so that $\Phit(\nt,\Jt)=b=\overline{4}\otimes 3\otimes \overline{1}\otimes
2\otimes 1\otimes 1$. The statistics in this case are
$E(b)=\cc(\nt,\Jt)=8$.
\end{example}

\subsection*{Acknowledgements} 
Many thanks to Professor T. Lulek and the Organizing Committee for 
the invitation to the Summer School at Myczkowce and for providing
excellent conditions for the meeting.
I would also like to thank the Max-Planck-Institut
f\"ur Mathematik in Bonn and the University of Wuppertal for hosting
me while this work was completed. This work was partially supported
by the Humboldt foundation and NSF grant DMS-0200774.

\end{document}